\documentclass[manuscript,screen,sigconf]{acmart}
\AtBeginDocument{%
  \providecommand\BibTeX{{%
    \normalfont B\kern-0.5em{\scshape i\kern-0.25em b}\kern-0.8em\TeX}}}

\setcopyright{acmcopyright}
\copyrightyear{2018}
\acmYear{2018}
\acmDOI{XXXXXXX.XXXXXXX}
\acmConference[Conference acronym 'XX]{Make sure to enter the correct
  conference title from your rights confirmation emai}{June 03--05,
  2018}{Woodstock, NY}
\acmBooktitle{PMBS '23: Performance Modeling, Benchmarking and Simulation of High Performance Computer Systems, SC23, Denver, Nov 13, 2023} 
\acmPrice{15.00}
\acmISBN{978-1-4503-XXXX-X/18/06}

\begin{document}

\title{Comparative evaluation of bandwidth-bound applications on the Intel Xeon CPU MAX Series}

\author{Istv\'an Z Reguly}
\email{reguly.istvan@itk.ppke.hu}
\orcid{0000-0002-4385-4204}
\affiliation{%
  \institution{P\'azm\'any P\'eter Catholic University, Faculty of Information Technology and Bionics}
  \streetaddress{Pr\'ater utca 50/a}
  \city{Budapest}
  \country{Hungary}
  \postcode{1083}
}
\renewcommand{\shortauthors}{Istvan Reguly}

\begin{abstract}
In this paper we explore the performance of Intel\textsuperscript{\textregistered} Xeon\textsuperscript{\textregistered} MAX CPU Series, representing the most significant new variation upon the classical CPU architecture since the Intel\textsuperscript{\textregistered} Xeon Phi\textsuperscript{\texttrademark} Processor. Given the availability of a large on-package high-bandwidth memory, the bandwidth-to-compute ratio has significantly shifted compared to other CPUs on the market. Since a large fraction of HPC workloads are sensitive to the available bandwidth, we explore how this architecture performs on a selection of HPC proxies and applications that are mostly sensitive to bandwidth, and how it compares to the previous 3rd generation Intel\textsuperscript{\textregistered} Xeon\textsuperscript{\textregistered} Scalable processors (codenamed Ice Lake)  and an AMD EPYC\textsuperscript{\texttrademark} 7003 Series Processor with 3D V-Cache Technology (codenamed Milan-X). We explore performance with different parallel implementations (MPI, MPI+OpenMP, MPI+SYCL), compiled with different compilers and flags, and executed with or without hyperthreading. We show how performance bottlenecks are shifted from bandwidth to communication latencies for some applications, and demonstrate speedups compared to the previous generation between 2.0x-4.3x.
\end{abstract}




\maketitle

\section{Introduction}

Over the past several decades, the advancement in technology has led to a remarkable increase in computational power, with this progression following an exponential curve \cite{denning2016exponential}. This trajectory, described by Moore's Law, has seen the number of transistors on a chip double approximately every two years, thereby significantly boosting the processing capabilities of microprocessors. Computation however, is just one side of the coin: data has to be moved to be computed upon. While the performance of memory systems both in capacity and bandwidth has also been growing at an exponential rate, but with a lower coefficient - leading to a growing gap between how many computations can be performed per second, and how much data can be moved per second. This can be characterized as the "flop/byte" ratio, which increased from around 1.0 in the 90s to well over a 100 on today's graphics processing units (GPUs).

In an attempt to mitigate this growing disparity, known as the "memory wall", recent technological innovations have focused on tighter integration between memory and processors. High Bandwidth Memory (HBM) \cite{jun2017hbm}, a high-speed RAM interface designed to be used in conjunction with compute-heavy architectures, has been a significant player in this arena. HBM allows for increased bandwidth by placing memory stacks closer to the processor, thereby reducing latency and power consumption. For a few years now, this technology has been incorporated in GPUs to maximize their memory-intensive operations. The Intel Xeon Phi Processors, a series of manycore CPUs (now discontinued), also leveraged HBM - the first to do so with an x86 architecture. Fujitsu\textsuperscript{\textregistered}'s A64FX integrated HBM memory with 48 ARM cores, becoming the building block for the Fugaku supercomputer which was \#1 on the Top500 list in 2020 and 2021. The latest development in this trend has been the introduction of HBM in the Intel Xeon CPU Max Series, marking another crucial milestone in the journey towards improving the balance between compute power and memory bandwidth. While previous CPU architectures with HBM utilized light-weight cores, the Xeon CPU MAX Series uses heavy-weight cores, representing a step up even in terms of single-thread performance compared to earlier Intel CPUs - making it a very interesting architecture for highly complex, yet fundamentally bandwidth-limited codes. 
 
High-performance computing (HPC) applications have a critical dependency on memory bandwidth, as many of these applications are essentially bandwidth-bound, as identified in a recent study \cite{9460517}. Traditional benchmarks, such as the High-Performance Linpack (HPL) \cite{hpl}, which has become an industry-standard ranking for HPC systems, are not fully representative of the memory bandwidth requirements in most real-world HPC applications. Recognizing the shift to being constrained by the speed of memory movement, there is an increasing focus on building HPC machines with an emphasis on memory movement rather than just computational capacity - with Japan's supercomputers (Earth Simulator, K computer, Fugaku) pioneering these efforts. 

As of July 2023, we could not find a single paper carrying out performance studies on the Intel Xeon CPU MAX Series - only a preprint focusing on a particular seismic application, showing a single data point \cite{jadhav4515781migration}, websites showing performance overviews \cite{Larabel}, and marketing materials. We therefore endeavor to give one of the first overviews and detailed studies of performance on the Xeon CPU MAX 9480, focusing on primarily bandwidth-bound codes where this platform can truly challenge the status quo.

\begin{figure*}[h]
    \centering
    \includegraphics[width=\textwidth]{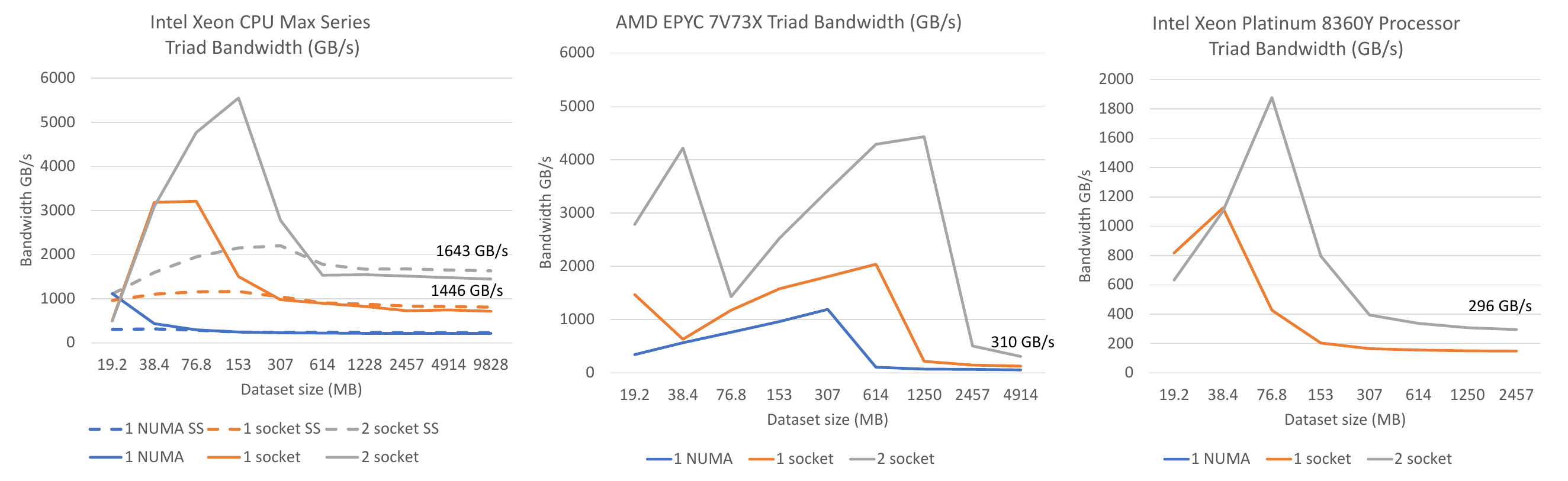} 
    \caption{BabelStream Triad bandwidth achieved on the three platforms}
    \label{fig:stream}
\end{figure*}

In this paper, we make the following contributions:
\begin{enumerate}
    \item We benchmark core-to-core latency and memory bandwidth on Intel Xeon CPU MAX 9480 Processor, as well as the Intel Xeon Platinum 8360Y Processor, and the AMD EPYC 7V73X processors, and contrast their raw performance figures,
    \item We explore the effects of choice of compiler, compiler flags (use of AVX512 instructions in particular), Hyperthreading, and parallelization approach on overall performance on the Intel Xeon CPU MAX Series running a range of applications,
    \item We carry out a detailed analysis of different parallelizations such as pure MPI, MPI+OpenMP, and MPI+SYCL, quantifying bottlenecks, and absolute performance metrics,
    \item We compare the best performance achieved on each application on the Intel Xeon CPU MAX 9480 Processor, the Intel Xeon Platinum 8360Y Processor, and the AMD EPYC 7V73X.
\end{enumerate}

The paper is structured as follows: Section \ref{sec/baseline} introduces the platforms, raw performance figures, and evaluates core-to-core latency and achieved bandwidth, Section \ref{sec/apps} introduces the applications benchmarked, Section \ref{sec/parallelizations} discusses the different parallelizations for each application, then Section \ref{sec/comparison1} gives an overview of compilers, configurations, and parallelizations. Section \ref{sec/comparison3} compares absolute performance metrics and to the other CPUs. Finally, Section \ref{sec/conclusions} draws conclusions.


\section{Systems and Baseline Performance}
\label{sec/baseline}
For this study, we evaluate the following systems, in the given hardware/software configuration.

\begin{enumerate}
    \item Intel Xeon CPU MAX 9480 Processor with HBM, available in the Intel Developer Cloud. Two sockets, each with 56 cores, Hyperthreading on. 2x4 NUMA regions, with 2x64 GB HBM in HBM-only mode, with SNC4. Clock frequencies between 1.9 GHz (base frequency) - 2.6 GHz (all-core turbo), giving a theoretical 13.6-18.6 FP32 TFLOPS/s. Software: Intel OneAPI Base and HPC toolkits, 2023.1 (including Intel MPI). Benchmarked Jun 4, 2023.
    \item Intel Xeon Platinum 8360Y Processor, available in the Baskerville cluster at the University of Birmingham. Two sockets, each with 36 cores, Hyperthreading on. 512 GB DDR4 RAM. Clock frequencies between 2.4 GHz (base frequency) - 2.8 GHz (all-core turbo), giving a theoretical 11-13 FP32 TFLOPS/s. Software: RHEL 8.5, Intel OneAPI Base and HPC toolkits, 2023.1 (including Intel MPI). Benchmarked July 11, 2023.
    \item AMD EPYC 7V73X with 3D V-Cache Technology, available as an Azure HB120rs\_v3 virtual machine. Two sockets, each with 60 available cores, Hyperthreading off. 2x2 NUMA regions, with 448 GB DDR4 RAM. Clock frequencies between 2.2 GHz (base frequency) - 3.5 GHz (turbo), giving a theoretical 8.45-13.45 FP32 TFLOPS/s. Software: Ubuntu 22.04, GCC 12.3 and AMD Optimizing C/C++ Compiler 4.0. Benchmarked July 15, 2023.
\end{enumerate}

\begin{figure*}[h]
    \centering
    \includegraphics[width=\textwidth]{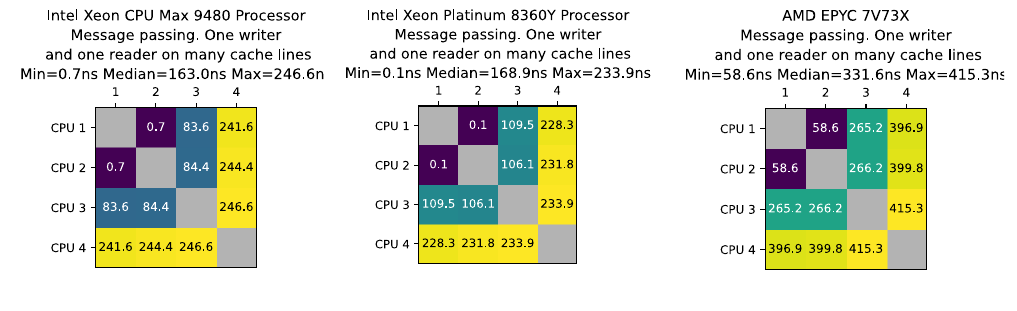} \vspace{-40pt}
    \caption{Message passing latency on the three platforms}
    \label{fig:latency}
\end{figure*}

The biggest highlight of the Intel Xeon CPU MAX Series is the massive improvement in memory bandwidth, thanks to the on-package high-bandwidth memory. While the theoretical peak bandwidth on Xeon 8360Y and EPYC 7V73X are 2x204.8 GB/s, on Intel Xeon CPU MAX Series this is around 2x1300 GB/s \cite{ixpug}. Therefore, we first benchmark the achievable memory bandwidth using the BabelSTREAM OpenMP Triad benchmark. Results are shown in Figure \ref{fig:stream} – measuring bandwidth from a single NUMA domain, one socket, or both sockets, at different array sizes. At the largest sizes, the Xeon Platinum 8360Y and the EPYC 7V73X achieve close to 75\% of peak at 296 GB/s and 310 GB/s respectively. In comparison, on the Intel Xeon CPU MAX Series we also evaluate two sets of compiler flags – one used on other platforms and in the applications, and one fine-tuned for the STREAM benchmark \cite{guide} (marked with SS for streaming stores). The former achieves 1446 GB/s, a 4.8x increase over the Xeon Platinum 8360Y and the EPYC 7V73X, the latter 1643 GB/s, a 5.5x increase over Xeon Platinum 8360Y/EPYC 7V73X. Still, only 55\%/63\% of peak is reached respectively – the reasons for this are explored in \cite{ixpug}. At smaller array sizes one can observe bandwidth to on-chip cache – what is particularly relevant here, is the ratio between cache and main memory bandwidth, which at 3.8x on Intel Xeon CPU MAX 9480 is significantly lower compared to 6x on the Xeon Platinum 8360Y and 14x on the EPYC 7V73X.

At base clock frequencies, the three platforms have a peak FP32 computational capacity of 13.6 TFLOPS/s (Intel Xeon CPU MAX 9480), 11 TFLOPS/s  (Xeon Platinum 8360Y), and 8.45 TFLOPS/s (EPYC 7V73X). The theoretical flop/byte ratio, indicating the balance between memory movement and compute is significantly reduced on the Intel Xeon CPU MAX 9480 Processor to 9.4, compared to 36 on the Xeon Platinum 8360Y and 28 on the EPYC 7V73X – indicating that bandwidth-bound applications on other platforms may become compute-bound (or latency-bound) on Intel Xeon CPU MAX Series. The flip side of this is that while the Intel Xeon CPU MAX 9480 Processor does have the highest computational capacity, it is only 24\% and 61\% higher compared to Xeon 8360Y and EPYC 7V73X respectively, therefore computationally bound applications are not expected to gain as much speedup compared to previous generation as bandwidth-bound ones.

Another important factor in HPC is the communication latency between CPU cores, because threads may need to synchronize, or processes may need to exchange data. To measure this, we use the core-to-core-latency test \cite{c2c} and evaluate message passing latency with the one writer/one reader on many cache lines test between (1) hyperthreads, (2) adjacent cores, and (3) cores on different sockets. While there were some noise in the measurements, Figure \ref{fig:latency} shows that  there hasn’t been a significant improvement (in some cases even slight regression) in communication latencies compared to the Xeon Platinum 8360Y – while this is not surprising given the lower clock frequency, higher core counts, and chiplet architecture, it is important to note as with the reduction of the bandwidth bottleneck, applications may become bound by this latency.
For EPYC 7V73X, since simultaneous multithreading was disabled, we show latencies to the adjacent core, a core in a different NUMA domain, but the same socket (different chiplet), and a core in the other socket. While overall latencies are comparable to the Xeon Platinum 8360Y and the Xeon CPU MAX 9480, the latency across different sockets is 1.6x times worse (note that this node was run as a virtual machine in Azure cloud, which might affect results).

\begin{figure*}
    \centering
    \includegraphics[width=0.99\textwidth]{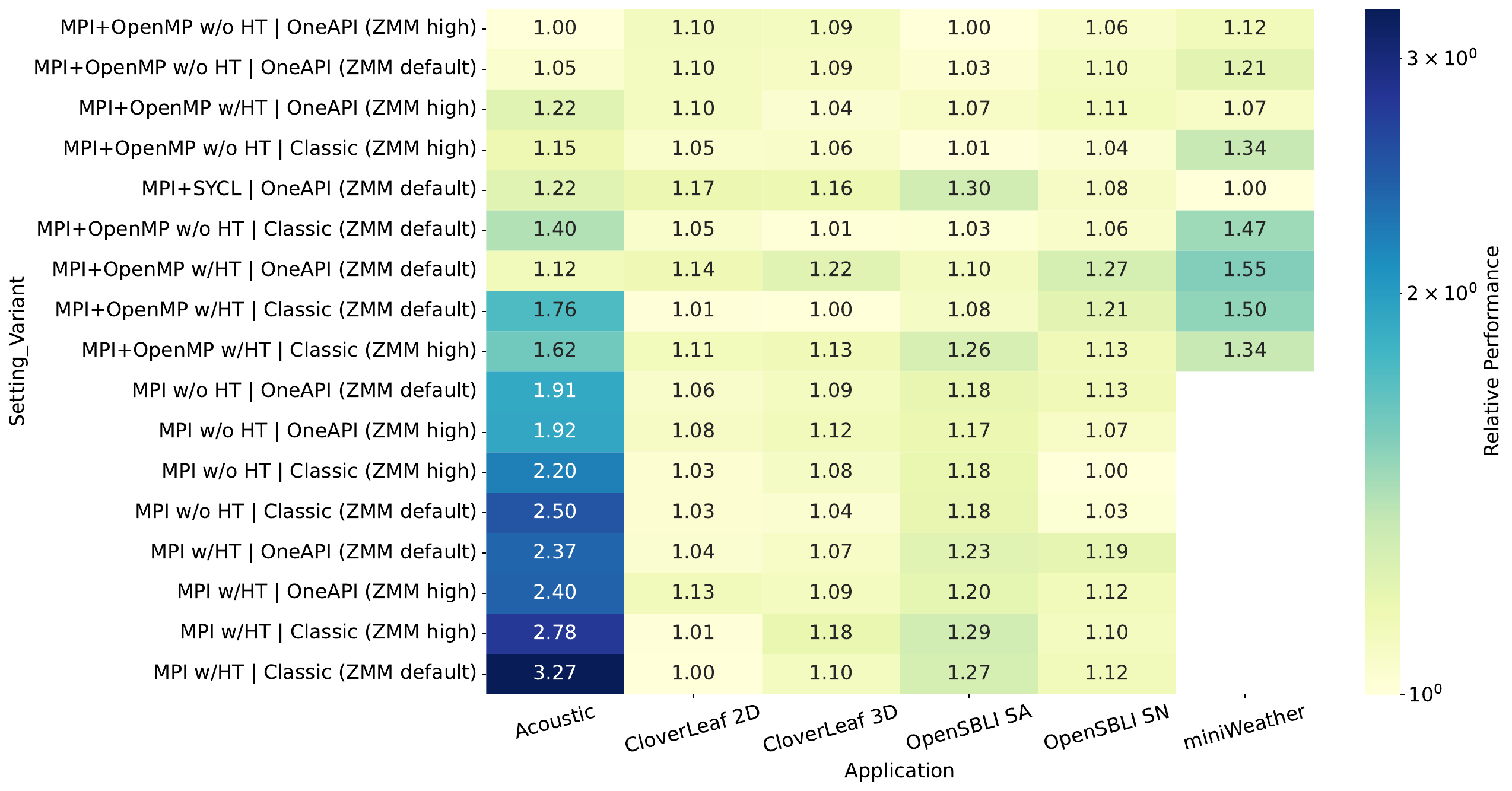} \vspace{-15pt}
    \caption{Configurations of structured mesh benchmarks normalized to the best for each application} \vspace{-10pt}
    \label{fig:structured}
\end{figure*}

\begin{figure}
    \centering
    \includegraphics[width=0.9\columnwidth]{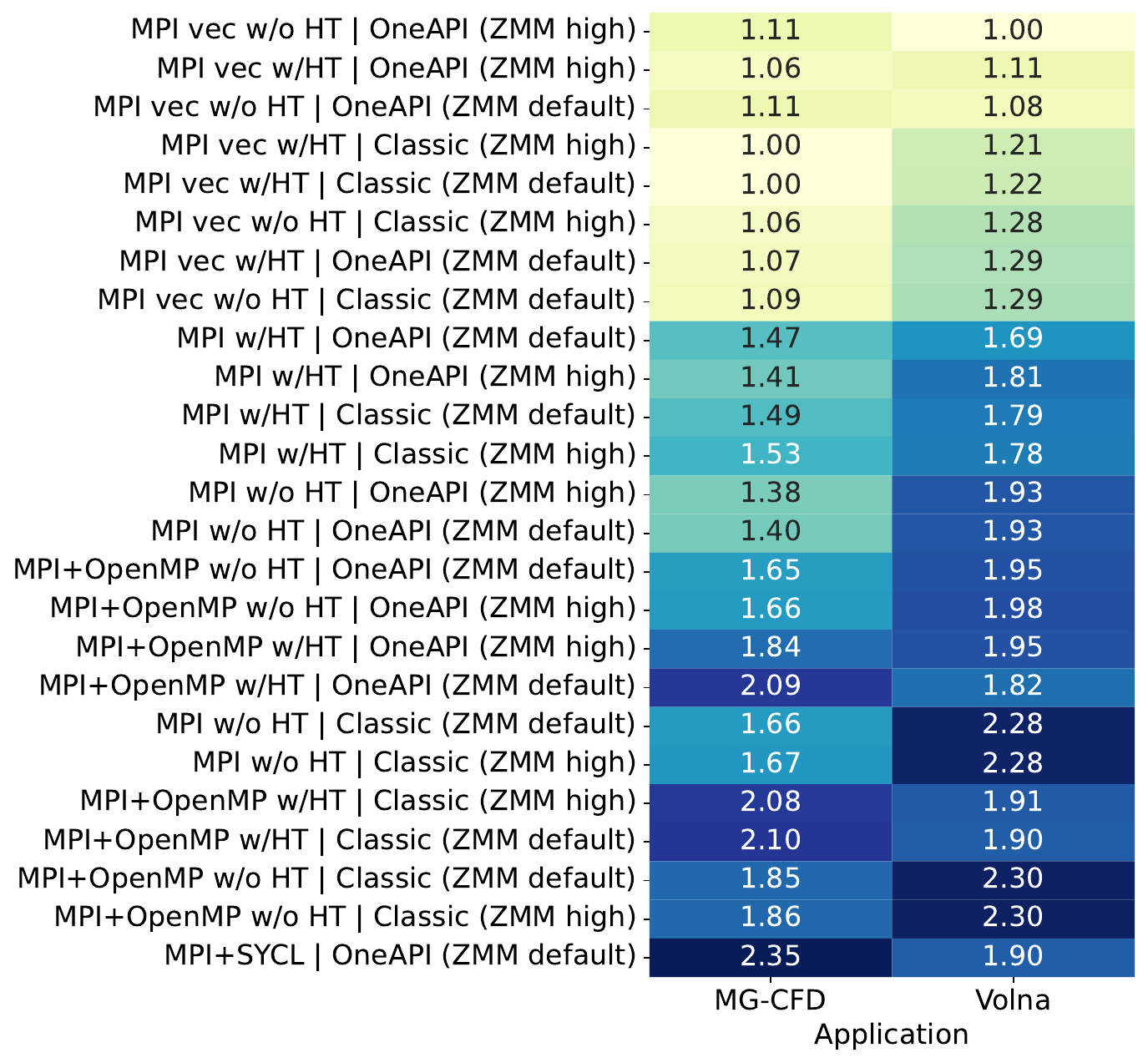} \vspace{-15pt}
    \caption{Configurations of unstructured mesh benchmarks normalized to the best for each application} \vspace{-10pt}
    \label{fig:unstructured}
\end{figure}

\section{Applications}

In the following, we briefly introduce the benchmarked applications and their key characteristics. All measurements were done 4 times, and averaged (variance in results was <5\%, and is not explicitly shown).

\label{sec/apps}
\begin{enumerate}
\item miniBUDE \cite{miniBUDE} – proxy molecular docking code, representative of BUDE. Compute and latency bound. Single precision, bm1 testcase, 30 iterations.
\item CloverLeaf 2D/3D \cite{cloverleaf} – structured-mesh Eulerian hydrodynamics simulation, representative of nuclear security codes. Mostly bandwidth-bound, with some operations on faces/edges that may be latency bound. Double precision, $7680^2$(2D), $408^3$(3D) problem size, 50 iterations.
\item Acoustic – structured-mesh high-order (8th) finite difference acoustic wave propagation solver. Bandwidth and cache locality bound, with large communications volume over MPI. Single precision, $320^3$ problem size, 10 time iterations.
\item OpenSBLI SA \& SN \cite{opensbli} – structured mesh finite difference Navier-Stokes solver for capturing shock-boundary layer interactions. Production code with 2 varians – Store All (SA), which is bandwidth-bound, and Store None (SN), which recomputes derivatives on the fly, reducing data movement pressure, but still mostly bandwidth bound. Double precision, $320^3$ problem size, 20 time iterations.
\item MG-CFD \cite{mgcfd} – unstructured mesh finite volume Euler equations solver with multigrid – proxy for Rolls-Royce’s CFD simulator Hydra. Bound by latencies and indirect memory accesses. Double precision, NASA Rotor37 case with 8 million vertices, 25 iterations.
\item Volna \cite{volna} – unstructured mesh finite volume Nonlinear Shallow Water Equations solver. Also sensitive to indirect memory accesses as MG-CFD, but less so. Single precision, Indian ocean case with 30 million vertices, 200 time iterations.
\item miniWeather \cite{miniweather} – structured mesh proxy code implementing basic dynamics seen in atmospheric weather and climate simulations. Bandwidth bound. Double precision, $4000x2000$ problem size, simulation time 1.0.
\end{enumerate}

\section{Parallelizations}
\label{sec/parallelizations}
All except two (miniWeather, miniBUDE) applications are implemented in the OPS \cite{ops} or OP2 \cite{op2} domain specific languages for structured/unstructured mesh applications. This enables these applications to automatically generate a variety of different parallel implementations, including MPI, MPI+OpenMP, MPI+SYCL and more. The performance of OPS/OP2 versions of applications have been thoroughly studied in prior work and shown to match hand-coded variants \cite{mudalige2015performance,mudalige2019large,7152942}.
For structured mesh computations (including miniWeather) a standard cartesian mesh decomposition is used over MPI, with ghost cell exchanges triggered as needed before each bulk parallel computational step. Within each process, OpenMP or SYCL parallelization may be applied – parallelizing across all grid points – using OpenMP’s collapse and simd pragmas, or in SYCL creating a work item for each grid point. For SYCL, there is an additional variation: ``flat'' uses a plain parallel\_for loop specifying the whole domain to be executed on, whereas ``ndrange'' uses a parallel\_for loop with an nd\_range argument, explicitly specifying workgroup shapes. Internally, the ``flat'' scheme is still mapped to workgroups by the runtime, however, the runtime may choose the exact shape for each kernel on each target architecture.
For unstructured mesh computations, we perform a standard owner-compute decomposition of the mesh over MPI using PT-Scotch \cite{chevalier2008pt}. Within each process for the MPI-only version we sequentially process elements, and can optionally use an explicitly auto-vectorizing implementation, where the generated code packs and unpacks vector registers. For OpenMP and SYCL one needs to explicitly avoid race conditions – for which we use a coloring scheme \cite{mgcfd-sycl}. While the OpenMP version does not auto-vectorize, we can generate SYCL code that vectorizes.

\section{Overview of applications, compilers, and configurations}
\label{sec/comparison1}

In this section, we explore the impact of basic configuration combinations on performance. Namely, we evaluate the various (feasible) combinations of the following:
\begin{enumerate}
    \item Compiler versions: Intel C++ Compiler \textbf{Classic} (ICC/ICPC) or Intel \textbf{oneAPI} DPC++/C++ Compiler (ICX/ICPX)
    \item ZMM usage set to default or high – affecting whether AVX-256 or AVX-512 instructions are generated. For bandwidth-bound applications it is not always clear which one is better, since the use of AVX-512 leads to fewer instructions, but lower clock speeds.
    \item Hyperthreading - 1 or 2 threads used per hardware core
    \item Parallelization approaches: (1) Pure MPI: assigning a process to each physical/logical core. No threading overheads, but significant message passing overheads. (2)	MPI + OpenMP: assigning one MPI process to each NUMA domain, and 1 OpenMP thread to each physical/logical core, (3) MPI + SYCL – assigning one MPI process to each NUMA domain, then using that as a SYCL sub-device.
\end{enumerate}

Overall results from structured mesh applications are shown in Figure \ref{fig:structured}, showing relative runtime (slowdown) vs. best for the given application, rows in ascending order by their average. As expected, the newer OneAPI compilers outperform the Classical compilers on average, although the absolute best performance is still achieved with the classical compilers on 3 out of 6 applications, with OneAPI within 4-6\%. On the other hand, for Acoustic the Classical compilers are 15\% slower, and for miniWeather 34\% slower.

MPI+OpenMP works best on average, performing best or only being 1\% slower for 4 applications – for OpenSBLI SN it is 4\% slower than pure MPI, and for miniWeather it is 7\% slower than MPI+SYCL. Hyperthreading disabled leads to marginally (2\%) better performance with the MPI+OpenMP codes, except for the CloverLeaf applications, but even there it’s within 3\% of the HT-enabled counterparts. The choice of ZMM usage does not have a substantial effect on these primarily bandwidth-bound codes – across all variants, ZMM default is only 0.2\% faster. It is only on the two most computationally intensive applications (Acoustic and OpenSBLI SN) is ZMM high consistently better (4-6\%).

Overall the best performing combination appears to be MPI+OpenMP, with OneAPI, ZMM high, and HT disabled. Yet this choice is still 6.2\% slower than picking the best option for each application individually.

Results from unstructured mesh applications are shown in Fig \ref{fig:unstructured}. Across both applications, MPI vec implementations (where auto-vectorizing code is generated for kernels with race conditions) perform the best – on average by 66\% compared to others. Here, ZMM high usage is required to match generated vector sizes, and the new oneAPI compilers work best for Volna, whereas the Classical compilers work better for MG-CFD. With critical kernels not vectorizing for either implementation, pure MPI variants are still on average faster than MPI+OpenMP due to the further loss in data locality. Here, the OneAPI compilers perform consistently better than the Classical compilers. For these cases, Hyperthreading enabled also improves performance by 13\% on average.

Finally, for the largely compute bound code miniBUDE, the Classical compilers generate code that stalls, therefore we could only measure with the OneAPI compilers. We achieve 6 TFLOPS/s with OneAPI, without HT and ZMM high. As a compute bound code, ZMM high improves performance by 45\% - however, one has to enable this explicitly. One thread per physical core can saturate pipelines – HT enabled reduces performance by 28\%. The SYCL implementation is not competitive, reaching only 50\% of OpenMP.

This analysis shows that the Xeon CPU MAX Series platform is particularly sensitive to the right configuration for bandwidth-limited codes. While we expect larger variations for the unstructured mesh applications, due to significant differences in data locality and vectorization between the different parallelizations, for the structured mesh codes there is much less inherent difference. The mean slowdown vs the best configuration on structured meshes is 1.25, with the median at 1.12. In comparison, the mean slowdown on the Xeon Platinum 8360Y is only 1.11, with the median at 1.05.

\subsection{Comparison of parallelizations on Intel Xeon CPU MAX Series}
\label{sec/comparison2}

\begin{figure}
    \centering
    \includegraphics[width=\columnwidth]{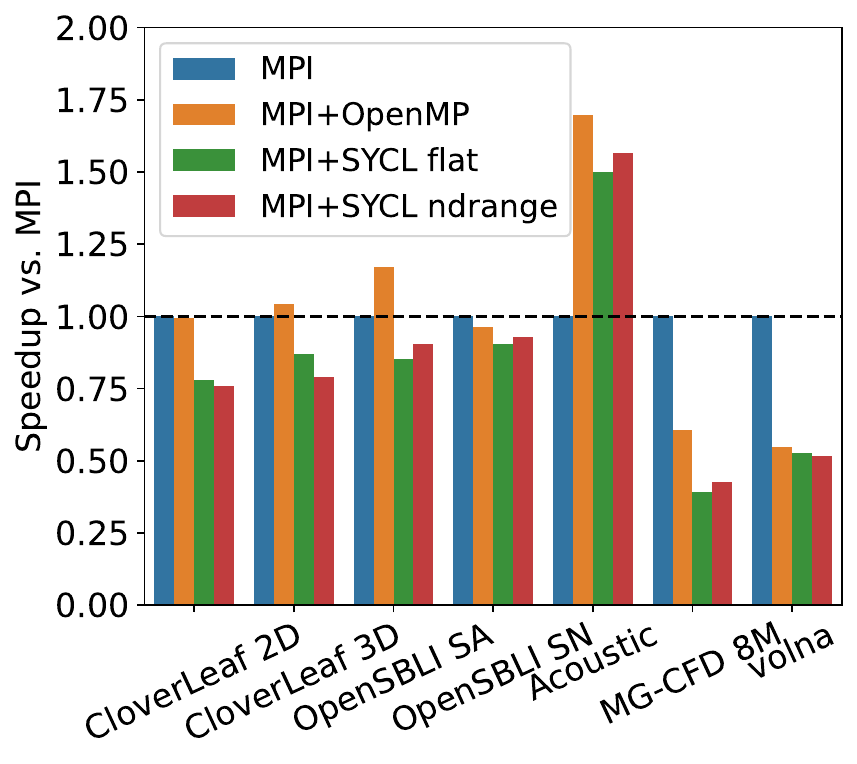} \vspace{-10pt}
    \caption{Relative speedup of different parallelizations on the Intel Xeon CPU MAX 9480 Processor compared to pure MPI}\vspace{-15pt}
    \label{fig:relative}
\end{figure}

\begin{figure*}
    \centering
    \includegraphics[width=\textwidth]{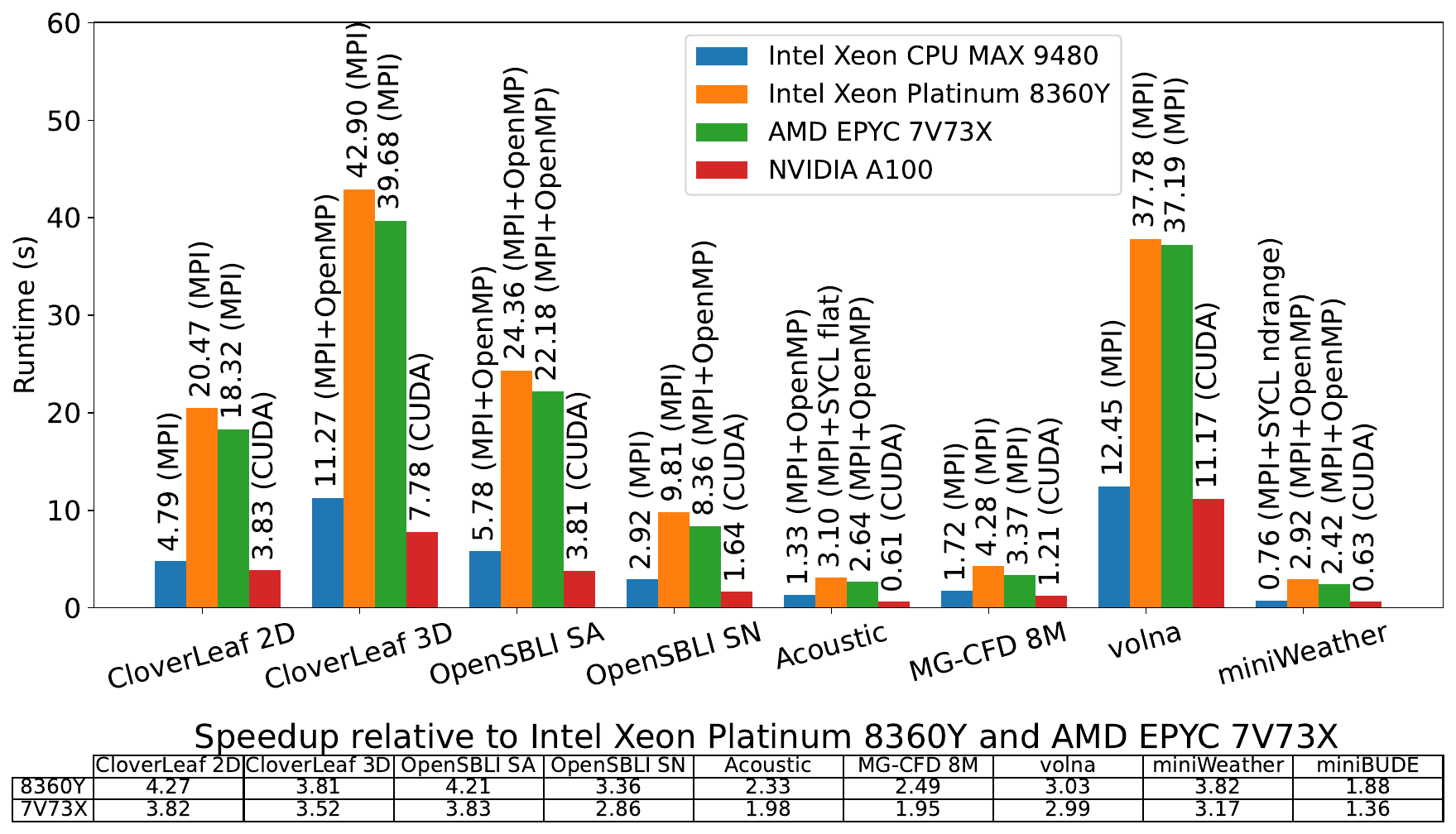} \vspace{-10pt}
    \caption{Best performance of applications on different platforms and the speedup of the Intel Xeon CPU MAX 9480 in comparison}
    \label{fig:comparison}
\end{figure*}
\begin{figure*}
    \centering
    \includegraphics[width=0.9\textwidth]{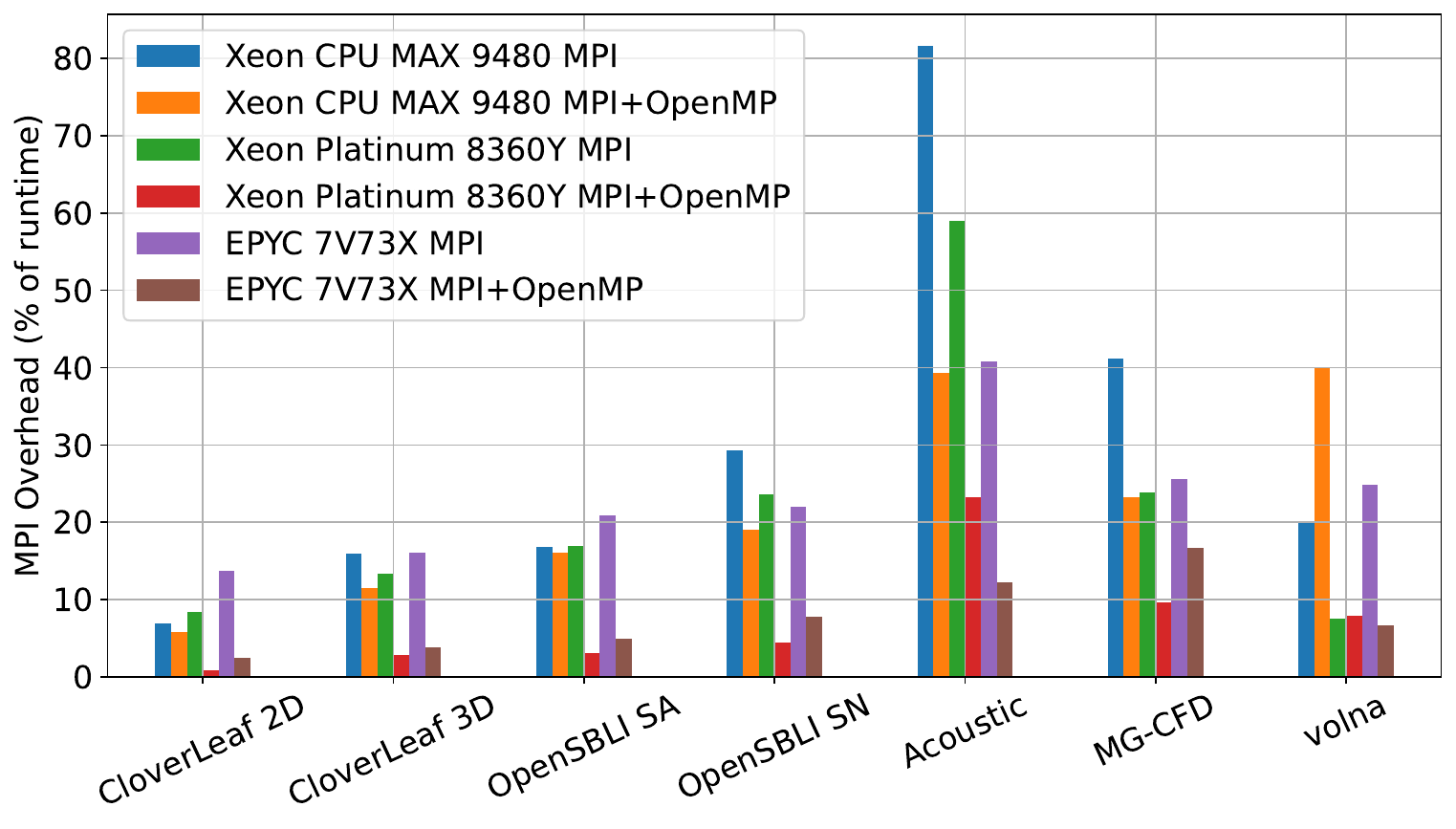} 
    \caption{Fraction of total runtime spent in MPI communications on different platforms for the different applications}
    \label{fig:mpi}
\end{figure*}

Figure \ref{fig:relative} shows that a hybrid MPI+OpenMP implementation (1 process per NUMA domain) performs best on average on structured mesh applications – especially if they are more limited by communications (Acoustic). On unstructured mesh applications (MG-CFD and volna), the MPI version auto-vectorizes, significantly outperforming (1.6-1.8x) MPI+OpenMP (which does not). MPI+SYCL at this point does not match the performance of MPI+OpenMP due to the higher scheduling overheads (having to go through the OpenCL drivers): this is more pronounced on CloverLeaf 2D/3D due to the higher number of small boundary kernels – less so on the more computationally intensive applications (OpenSBLI SN, Acoustic, volna). On MG-CFD, while the SYCL version does vectorize (flat version), it is still slower than the ndrange version with 1 work item (which disables vectorization) - reasons for this are explored in \cite{mgcfd-sycl}.
SYCL has two variants for structured meshes: “flat”, where the runtime chooses workgroup sizes for each kernel, and “ndrange”, where the user specifies one workgroup size to be used for all kernels in a given application. The latter we fine-tuned through exhaustive search, but we can see that at the level of the whole application, the runtime does a very good job at picking good workgroup sizes. Looking at an individual kernel for OpenSBLI SN, we can observe that better performance is achieved when the workgroup size in the contiguous dimension matches the size of the domain, and the other dimensions are small – in this case a shape of 160x4x4 gave 2\% faster execution than the default size with “flat”. This is consistent with our understanding of cache prefetchers and task granularity – for GPUs of course this would be different, where the maximum workgroup size is 1024. This highlights the importance of choosing workgroup sizes specific to a kernel and a target architecture, which the runtime – given the right heuristics – is better equipped to do compared to the user hard-coding these parameters.

\section{Comparison to other platforms}
\label{sec/comparison3}

Figure \ref{fig:comparison} shows a performance comparison of the Intel Xeon CPU MAX Series platform to the previous-generation Xeon 8360Y CPU, the EPYC 7V73X CPU, and an NVIDIA A100 (40GB PCI-e), with the  data labels indicating the best performing implementation on the given hardware, and a table underneath showing speedups on Intel Xeon CPU MAX 9480 vs. Xeon 8360Y and EPYC 7V73X. The most bandwidth-bound codes (CloverLeaf 2D and OpenSBLI SA achieve a speedup of 4.2/3.8x respectively – slightly less than the difference in peak bandwidth (4.8x), due to a lower bandwidth utilization efficiency. 
Communications latency and cache performance is a more significant factor in the overall performance of OpenSBLI SN and Acoustic, but speedups are still over 2.5/1.98x. It is interesting to note that the OpenSBLI SA and SN are different formulations of solving the same problem, with SN doing more compute and SA moving more data – the speedup between these two is just below 2x on Intel Xeon CPU MAX 9480, but over 2.5x on Xeon 8360Y/EPYC 7V73X, showing that trading off data movement for computations is still worthwhile, but less effective than on more bandwidth-constrained platforms. For the MG-CFD unstructured mesh applications there is an overhead of packing and unpacking vector registers – since EPYC 7V73X only has 256-bit AVX2, this overhead is smaller, and thanks to its large cache, locality is significantly improved compared to Xeon 8360Y – leading to lower speedups on Intel Xeon CPU MAX 9480 at 2.5/2x respectively. Finally, miniBUDE is much more bound by latency and compute, but there is still significant speedup thanks to the availability of AVX512 on Intel Xeon CPU MAX 9480 – 1.9/1.36x respectively.

GPUs, generally speaking, excel at streaming computations such as these (the structured mesh ones in particular) thanks to their high memory bandwidth, but of course the Xeon CPU MAX Series comes equipped with a similar memory technology - and so we take a brief look at comparing performance to and NVIDIA A100 GPU - which has an achievable peak memory bandwidth of 1310 GB/s - 10\% lower than that measured on the Intel Xeon CPU MAX 9480. Yet, as shown in Figure \ref{fig:comparison} the A100 is significantly (1.1-2.1$\times$) faster. The difference is notably lower on more bandwidth-bound codes, but more pronounced on the more compute-bound codes such as OpenSBLI SN and Acoustic. The difference in part comes down partly to better bandwidth utilization (thanks to the massive SMT capabilities of GPUs), and no MPI communications overheads.

As previously discussed, the communication latencies on Intel Xeon CPU MAX Series did not improve by as much as bandwidth did, resulting in a shift of bottlenecks. In Figure \ref{fig:mpi} we quantify this overhead by measuring the time spent in MPI\_Wait for different applications on the three hardware platforms with pure MPI and MPI+OpenMP implementations. The first clear difference is that for all but one application (volna), the MPI+OpenMP implementation has significantly lower MPI overhead - which is as expected, given that fewer messages are being sent and the overall communications volume is smaller as well. As shown in Figure \ref{fig:comparison}, this does not mean that MPI+OpenMP performs best overall - notably for CloverLeaf 2D/3D and OpenSBLI SN pure MPI runs faster - OpenMP has its own overheads with the need for synchronization and data sharing between threads (neither applies to pure MPI). Whereas the average improvement in communications overheads when going from MPI to MPI+OpenMP for the Xeon Platinum 8360Y and the EPYC 7V73X are at 15\%, for the Xeon CPU MAX 9480 this is only 8.2\% - once again underlining the shift from the bandwidth bottleneck to the latency bottleneck.
Aside from CloverLeaf 2D, we can see that the percentage of time spent in MPI on Intel Xeon CPU MAX 9480 is 1.2-5.3x higher compared to Xeon Platinum 8360Y – of course in absolute terms MPI is still only 0.8x slower on OpenSBLI SA and 2.7x faster on OpenSBLI SN. 

\begin{figure}
    \centering
    \includegraphics[width=\columnwidth]{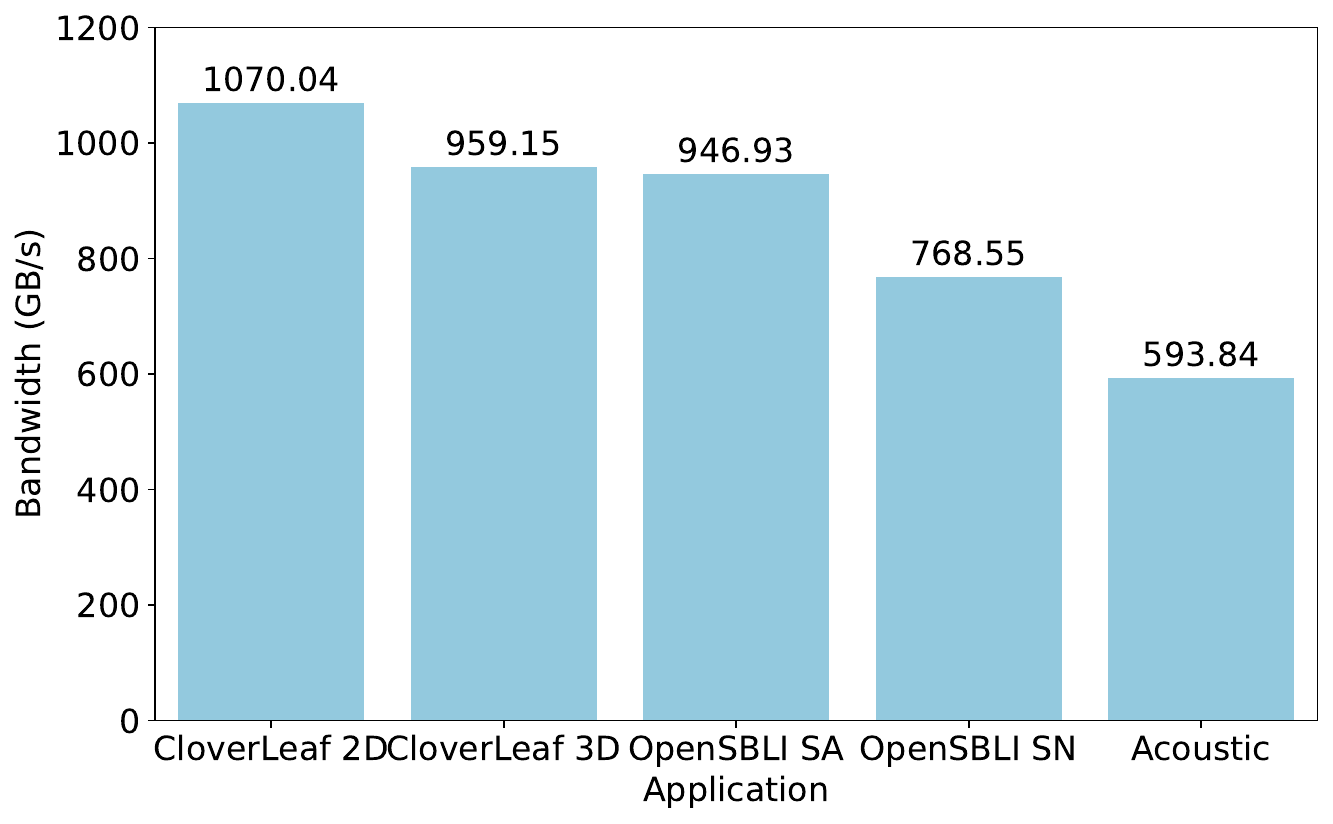}  \vspace{-20pt}
    \caption{Achieved effective bandwidth of applications on the  Intel Xeon CPU MAX 9480 Processor}\vspace{-15pt}
    \label{fig:achieved_bw}
\end{figure}

\begin{figure}
    \centering
    \includegraphics[width=\columnwidth]{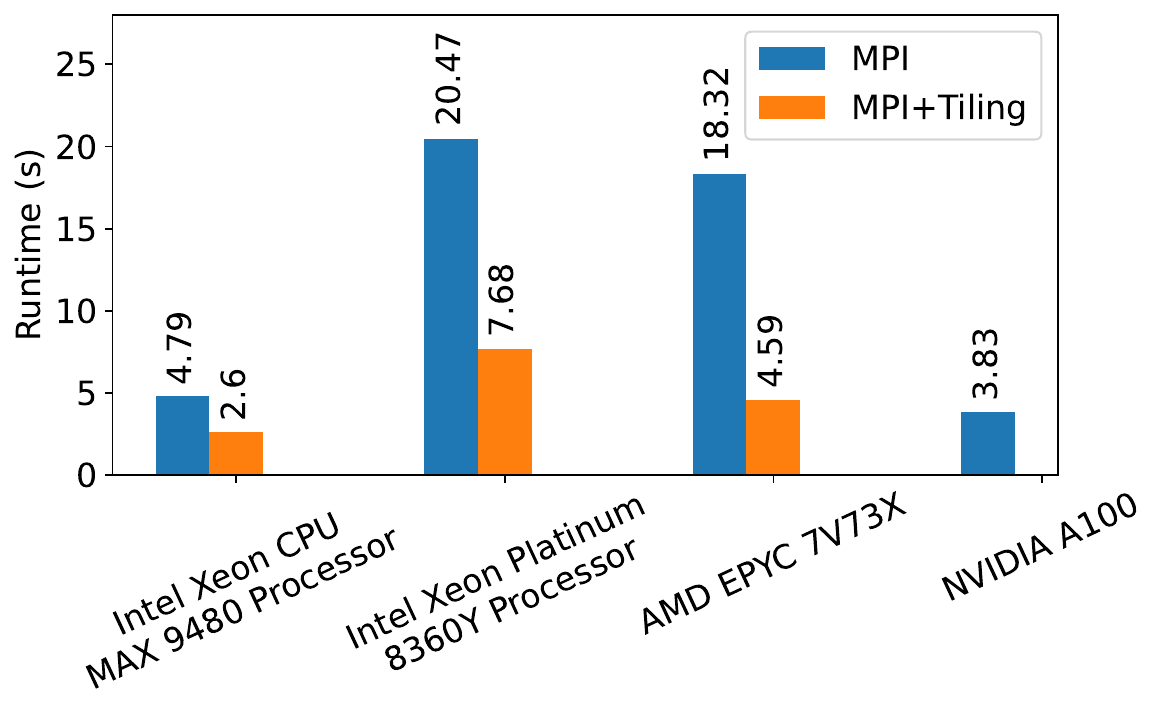} 
    \caption{CloverLeaf 2D with the cache-blocking tiling optimization - comparison of different platforms}
    \label{fig:tiling}
\end{figure}

We also evaluated absolute performance metrics – namely achieved effective memory bandwidth of the kernels. This was calculated by OPS automatically, by measuring the execution time of the kernel (excluding MPI communications), and estimating the effective data movement, based on the iteration ranges, datasets accessed, and types of access (read or read+write). CloverLeaf 2D achieves 75\% of peak – it is the application with the simplest access patterns, and low data re-use. The CloverLeaf 3D and OpenSBLI SA applications are also memory intensive, but given they are in 3D, their access patterns are more complicated – they still achieve over 65\% of peak. OpenSBLI SN and Acoustic are more computationally intensive, and Acoustic in particular is also very cache-intensive due to the high-order stencils - these achieve only 53\% and 41\% of peak respectively. In comparison, Xeon 8360Y achieves 75-85\% of peak and EPYC 7V73X achieves 79-96\% of peak on these applications, showing that the bandwidth bottleneck on Intel Xeon CPU MAX Series is significantly reduced.

Finally, we take the still most bandwidth-bound code, CloverLeaf 2D, and apply the cache-blocking tiling algorithm in OPS. This algorithm \cite{reguly2017loop} re-arranges the execution of parallel loops within and across different loops to improve memory locality. It also reduces MPI communications volume and frequency, at the cost of redundant computations along the MPI boundaries.  Here we used the OneAPI compilers with ZMM usage high, running pure MPI with hyperthreading on Intel, and AMD Optimizing C/C++ Compiler (AOCC) on the AMD EPYC 7V73X. Figure \ref{fig:tiling} shows the improvements in runtime – for reference we also added the performance of CUDA running on an A100 40GB GPU as well. We can see that performance on the Intel Xeon CPU MAX 9480 CPU can be improved by a further 1.84x by increasing cache re-use, at this point outperforming an A100 GPU by 1.5x. It is also interesting to note that the same improves performance by 2.7x on the Xeon Platinum 8360Y and 4x on the EPYC 7V73X, a significantly higher factor, but it correlates well with the difference between measured cache bandwidth and HBM/DDR4, which is 3.8x on Intel Xeon CPU MAX 9480, 6.3x on the Xeon Platinum 8360Y, and 14x on the EPYC 7V73X.

\section{Conclusions}
\label{sec/conclusions}

In this paper we have conducted a qualitative and quantitative benchmarking of the Intel Xeon CPU MAX 9480 Processor - with a particular focus on traditionally bandwidth-bound HPC applications. After measuring basic achievable bandwidth and message-passing latency figures, we explored how the choice of compiler, compiler flags, parallelization approaches, and hyperthreading affects performance, showing a significantly higher variation compared to previous generations. Having quantified the reduction in the bandwidth bottleneck, but only moderate improvements in terms of communication latencies, we contrasted the performance of pure MPI, MPI+OpenMP and MPI+SYCL implementations of the same codes, trading off communications overheads for threading overheads, showing improvements in the communications overheads, but still overall a mixed picture in terms of which implementation performs best. We then proceeded to calculate the effective bandwidth utilization, reaching 41-75\% of peak (as measured by BabelStream). 
The comparison with the previous-generation Xeon Platinum 8360Y and the AMD EPYC 7V73X clearly shows the shift away from the bandwidth bottleneck - revealing the importance of minimizing communications and best utilizing the vector units. Despite a lower architectural efficiency (i.e. lower fraction of peak bandwidth), the Xeon CPU MAX 9480 still outperformed the other CPUs studied by 2-4.3$\times$ on these benchmarks, suggesting that it could offer significant benefits compared to traditional DDR-only systems. With the right optimizations on bandwidth-bound applications (CloverLeaf 2D with cache-blocking tiling), the Xeon CPU MAX Series is even capable of outperforming modern GPUs - an NVIDIA A100 40GB PCI-e by 50\% in this case.

\section{Acknowledgments}

We are grateful for the support of the OneAPI Innovator program, and the advice and assistance of Mark Lubin, Xiao Zhu, and Rob Muller-Albrecht at Intel in particular. 

This research was supported by Rolls-Royce plc., and by the UK EPSRC (EP/S005072/1 – Strategic Partnership in Computational Science for Advanced Simulation and Modelling of Engineering Systems – ASiMoV). This work was also supported in part by the Hungarian Academy of Sciences under Grant POST-COVID2021-64.

\bibliographystyle{ACM-Reference-Format}
\bibliography{sample-authordraft}

\end{document}